\documentclass[10pt,conference]{IEEEtran}
\usepackage[pdftex]{graphicx}
\usepackage[usenames,dvipsnames,table,xcdraw]{xcolor}
\usepackage{array}
\usepackage{multirow}
\usepackage{siunitx}
\usepackage[normalem]{ulem}
\useunder{\uline}{\ul}{}
\usepackage{booktabs}
\usepackage{supertabular,booktabs}
\usepackage{longtable}
\usepackage{lscape}
\usepackage{cite}
\usepackage{url}
\usepackage{dirtytalk}
\usepackage{graphicx}
\usepackage{comment}
\usepackage{makecell}
\usepackage{svg}
\usepackage{amsmath}
\usepackage{balance}
\usepackage[caption=false]{subfig}
\graphicspath{{img/}}
\usepackage[normalem]{ulem}
\usepackage{arydshln}
\useunder{\uline}{\ul}{}

\usepackage{tabularx}
\usepackage{enumitem}
\usepackage{calc}
\usepackage{parskip}
\usepackage[para,online,flushleft]{threeparttable}
\usepackage[most]{tcolorbox}
\usepackage{nameref}

\newtcolorbox{mytextbox}[1][]{%
  sharp corners,
  enhanced,
  colback=white,
  attach title to upper,
  #1
}

\begin{document}

\title{User Personas Improve Social Sustainability by Encouraging Software Developers to Deprioritize Antisocial Features}

\author{\IEEEauthorblockN{Bimpe Ayoola, Miikka Kuutila, Rina R. Wehbe and Paul Ralph}
\IEEEauthorblockA{\textit{Faculty of Computer Science},
\textit{Dalhousie University},
Halifax, NS, Canada \\
bimpe.ayoola@dal.ca, miikka.kuutila@dal.ca, rina.wehbe@dal.ca, paulralph@dal.ca}
}


\IEEEtitleabstractindextext{%
\begin{abstract}
\textit{Background}: Sustainable software development involves creating software in a manner that meets present goals without undermining our ability to meet future goals. In a software engineering context, sustainability has at least four dimensions: ecological, economic, social, and technical. No interventions for improving social sustainability in software engineering have been tested in rigorous lab-based experiments, and little evidence-based guidance is available. 
\textit{Objective}:  The purpose of this study is to evaluate the effectiveness of two interventions---stakeholder maps and persona models---for improving social sustainability through software feature prioritization.
\textit{Method}: We conducted a randomized controlled factorial experiment with 79 undergraduate computer science students. Participants were randomly assigned to one of four groups and asked to prioritize a backlog of prosocial, neutral, and antisocial user stories for a shopping mall's digital screen display and facial recognition software. Participants received either persona models, a stakeholder map, both, or neither. We compared the differences in prioritization levels assigned to prosocial and antisocial user stories using Cumulative Link Mixed Model regression. 
\textit{Results}: Participants who received persona models gave significantly lower priorities to antisocial user stories but no significant difference was evident for prosocial user stories. The effects of the stakeholder map were not significant. The interaction effects were not significant. 
\textit{Conclusion}: Providing aspiring software professionals with well-crafted persona models causes them to de-prioritize antisocial software features. The impact of persona modelling on sustainable software development therefore warrants further study with more experience professionals. Moreover, the novel methodological strategy of assessing social sustainability behavior through backlog prioritization appears feasible in lab-based settings. 
\end{abstract}

}

\maketitle

\IEEEdisplaynontitleabstractindextext

\IEEEpeerreviewmaketitle

\section{Introduction}\label{sec:introduction}

Sustainable Development refers to meeting the needs of the present without compromising the ability of future generations to meet their own needs~\cite{brundtland1987report}. Sustainable software engineering, analogously refers to creating and maintaining software in a way that meets present needs without undermining our collective capacity to meet our future needs~\cite{Sean_sust}. Sustainable software development therefore involves minimizing negative impacts on the economy, society, human beings, and the environment during software development, deployment, and usage~\cite{naumann2011greensoft}. Put another way, sustainable software development means allowing technology to improve our daily lives while mitigating social risks (e.g. cybercrime, surveillance capitalism, social isolation)~\cite{al2014social}. 

Software sustainability encompasses four dimensions: \textit{economic} sustainability concerns wealth creation and profitability; \textit{environmental} (or \textit{ecological}) sustainability relates to the responsible use of natural resources; \textit{technical} sustainability involves designing quality software for adaptability~\cite{penzenstadler2013generic}; \textit{social} sustainability relates to impacts on the wellbeing of individuals, organizations, communities, and society~\cite{Sean_sust}. The social dimensions is the most complex and difficult to define of the four. 

Sustainability is rarely addressed holistically~\cite{gustavsson2020blinded}. Researchers and practitioners often focus on environmental or technical sustainability of software~\cite{garcia2017environmental, anwar2017, wolfram2017sustainability, moises2018practices, mourao2018green, CALERO2017117, duboc2019we}, but rarely address social sustainability despite sometimes being considered as important~\cite{gustavsson2020blinded, Sean_sust}. This gap exists partly because social sustainability is qualitatively complex and lacks clear criteria, practical guidance, or tools for integrating it into software development processes~\cite{Sean_sust}. For instance, a software development team may address environmental sustainability by enhancing energy efficiency,  technical sustainability through refactoring, and economic sustainability through market research and user-centered design (emphasizing a feasible revenue model). 

However, integrating social sustainability is more challenging. Development is socially sustainable when it meets present needs while improving, or at least not undermining, human wellbeing. Social sustainability involves creating software that benefits society, intersecting social issues such as equity, social justice, privacy, security, human well-being~\cite{chitchyan2016sustainability}. It takes different forms at different levels; for example, individual physical and mental health; group cohesion; organizational inclusiveness, community resilience, national social justice, or world peace. Therefore, development may be unsustainable in many ways; for instance, by deleteriously affecting safety, equity, tolerance, democratic participation, human rights, the rule of law, education, or sustainability awareness. 

Socially sustainable software development requires not only understanding the societal impact of software products  but also embedding principles that support social well-being and ethical responsibilities into the development process. This aspect of sustainability, often overshadowed by its environmental and technical counterparts, demands a more nuanced approach---delving into social values and impacts such as equality, diversity, community building, and belongingness. 

The challenges in incorporating sustainability into software development are further highlighted by a recent meta-analysis of sustainable software engineering research~\cite{Sean_sust}. This analysis revealed a critical gap: only a few studies have evaluated sustainability-promoting interventions~\cite{penzenstadler2014systematic,mourao2018green,marimuthu2017software,garcia2018interactions}, and none of these studies used controlled experiments. While controlled experiments are not the \textit{only} way to evaluate an intervention, the absence of any controlled experiments is a red flag. This underscores the need for more research to develop effective strategies for integrating social sustainability in software. 

This study begins to address this gap by evaluating the role of stakeholder mapping and personas in contributing towards social sustainability. WE pose the following research question. 

\begin{quote}
    \textit{\textbf{Research Question:} To what extent do stakeholder maps and personas affect how software professionals’ prioritize social sustainability?}
\end{quote}

Stakeholder mapping is the process of identifying and categorizing stakeholders by considering their needs, their concerns, and how a product or project might affect them~\cite{penzenstadler2013advocate, majumdar2013thorny}. Personas, on the other hand, are fictional characters created based on research to represent different user types. They help developers represent intended users to understand their needs, goals, characteristics and diversity~\cite{nielsen2019personas,faily2013guidelines,salminen2021survey}. They can also be used as a tool to facilitate sustainable user behaviour in software design~\cite{siddall2011personas}. 

\section{Related Work} \label{sec:RelatedWork}

\subsection{Social Sustainability} \label{sec:ss}

Social sustainability focuses on maintaining social capital and preserving the solidarity of communities~\cite{penzenstadler2013generic}, as well as ensuring equal or improved access to social resources for future generations, highlighting the importance of generational equity~\cite{lago2015framing, lago2021designing}. This concept is critical in the context of software development, where socially sustainable software should aim for digital inclusion by eliminating barriers, promoting multicultural awareness, and making systems more accessible to a diverse user base. This includes improving accessibility for the elderly, people with disabilities, non-English speakers, and low-literacy users to provide equal opportunities to the users~\cite{albertao2010measuring}.

However, accounting for the social dimension of sustainability is challenging for practitioners~\cite{al2016engineering} due to its intangible, qualitative nature and lack of consensus on relevant criteria~\cite{von2006accounting,mckenzie2009adult}. There is no clear method or comprehensive set of metrics for evaluating the social sustainability of a software product~\cite{al2014quantification}. Existing guidelines assessing social sustainability are often vague and inadequate~\cite{hajai}. In a recent survey, zero of 178 software professionals reported considering social sustainability in their development process or being guided by an explicit sustainability policy~\cite{oyedeji2024integrating}.

Significant attention has been directed towards the societal impacts of technology, particularly how AI technologies can affect marginalized and vulnerable populations~\cite{buolamwini2018gender, sin2021digital}. The increasing integration of AI in software systems brings unique challenges to social sustainability. An exploratory study~\cite{hajai} on the negative social impacts of existing AI systems highlights concerns such as discrimination, effects on people with disabilities, socio-economic division, and inequality in opportunities. Issues range from discriminatory AI in health predictions to a variety of threats including malicious use, trust erosion in authorities, and gender inequality. The study also emphasizes the lack of user awareness regarding AI discrimination and underscores the need for developing guidelines to promote socially sustainable AI technologies. 
In addressing these challenges, the role of human-centered design (HCD) practices becomes paramount, especially those that promote social justice. Traditional HCD approaches often center the experiences of dominant groups, inadvertently pushing vulnerable populations behind~\cite{rose2018social}. There is a need to shift towards more inclusive design practices, and implementing a social justice framework to prevent  marginalization and promote inclusive design\cite{rose2018social}.

Anticipatory measures, which involve proactive efforts to predict and prepare for potential future developments and risks associated with innovations, are also increasingly recognized as essential in technology governance. Stilgoe’s framework for Responsible Innovation, which identifies anticipation, inclusion, reflexivity, and responsiveness as its core dimensions, emphasizes the importance of foresight in understanding and addressing the societal implications of emerging technologies~\cite{stilgoe2020developing}. Anticipation, in particular, encourages stakeholders to consider the long-term societal implications of technological advancements. The Responsible Innovation framework further highlights that technology and society are deeply interconnected, and as such, innovation must align with societal values and ethical standards. Embedding anticipatory practices not only addresses immediate concerns but also anticipates future societal impacts, ensuring that technological advancements contribute positively to society~\cite{steen2021responsible}.

Al Hinai and Chitchyan~\cite{al2014social} identified over 600 social sustainability indicators in their systematic literature review, grouping them into 12 categories including including ``employment'', ``health'', ``education'', ``human rights'', and ``equality''. Their subsequent work~\cite{al2015building} focused on equality, suggesting that recognizing value patterns can help software engineers document the social sustainability requirements of the software under development. 

Most recently, Moises de Souza et al.~\cite{moises2023social} conducted a systematic literature review identifying various social aspects of sustainability in software development. These include ``diversity'', ``trust'', ``equality'', ``security'', ``transparency'', ``human rights'', and ``fairness''. They highlight the need for software practitioners to adopt practices that promote these social goals, such as fostering inclusive environments, upholding ethical standards, protecting human rights, and promoting transparency and fairness. The review also identified tools and practices aimed at fostering awareness of societal implications and ensuring alignment of software development with social sustainability goals, although they noted that these tools require further empirical validation to ensure their practicality and effectiveness in real-world software development scenarios. Tools like the Sustainability Awareness Framework~\cite{duboc2019we} aim to raise software engineers' awareness of the impacts of their work on societal sustainability. However, the effectiveness of these tools and practices still requires further empirical validation. McGuire et al.~\cite{Sean_sust} advocate for more research  evaluating the effectiveness of sustainability interventions.

\subsection{Stakeholder Mapping} \label{sec:sm}
An organization's \textit{stakeholders} are individuals or groups who can affect or are affected by its objectives or achievements~\cite{pacheco2012systematic}. Analagously, a project's stakeholders are individuals or groups who can affect or are affected by the project and its outcomes. \textit{Stakeholder mapping} is the process of visualizing an organization or project's  stakeholders; the visualization is called a \textit{stakeholder map}. Stakeholder maps may also include information about stakeholder's values, interests, relationships to one another, or how the product might affect them~\cite{majumdar2013thorny}. 

Johann et al.~\cite{johann2013social} emphasize the pivotal role of user and community engagement in fostering sustainable software engineering. Their work underscores the significance of socio-cultural contexts and promotes including diverse user backgrounds throughout the development process. This perspective resonates with the broader understanding that software products relies on several stakeholders including end-users. These stakeholders are not always fully identified or considered during software development and often times, only the most obvious or prominent ones are taken into account~\cite{majumdar2013thorny, wheeler2003focusing}. 

Several studies have identified stakeholder identification as a tool that promote sustainability~\cite{penzenstadler2013advocate,majumdar2013thorny,becker2015requirements,saputri2021integrated}. Johann et al.~\cite{johann2013social} also pointed out the importance of thinking about where users come from and their backgrounds when making software products, ensuring everyone is included. However, while these tools and methods sound great for making software more sustainable, there is still more work to do in validating and assessing them in a practical software development environment.

\subsection{Personas} \label{sec:p}

Personas are fictional characters used to describe user characteristics, goals, attitudes, and experiences when using software products~\cite{Cooper1999}. Personas go beyond mere statistical representations, descriptions, or basic information of average users; they weave together actual user behavior patterns, motivations, and frustrations with fictional information such as names, images, and backgrounds thereby rendering them more tangible and relatable~\cite{huynh2021building}. 

Personas enhancing the cognitive processes of designers, and foster a user-centric approach to design~\cite{dahiya2021effect}. Without the guidance of personas, designers tend to over-rely on their personal experiences or assumptions. Personas help anchor product design in the actual needs of users, mitigating the risk of poorly informed design choices and subjective judgments by software engineers~\cite{karolita2023use}. 

Creating accurate personas is time-consuming~\cite{hill2017gender}, and several methodologies have been proposed~\cite{grudin2002personas, pruitt2003personas, jansen2022create}. Some of these studies have also offered insights into effective persona integration in requirement engineering processes~\cite{karolita2023use}, and agile methodologies~\cite{losana2021systematic}. 

Despite their benefits, many practitioners still struggle to adopt personas, view personas with skepticism, or perceive them as too abstract, misleading, or irrelevant~\cite{chapman2006personas, matthews2012designers}. This could be due to a lack of familiarity among software practitioners or the absence of validated studies or practical guidelines for using personas effectively~\cite{karolita2023use}. This gap underscores the need for further research evaluating and validating personas with software practitioners to enhance the understanding of user needs in software engineering.

\section{Method} \label{sec:method}

To address our research question, we carried out a randomized, controlled factorial experiment with 79 undergraduate computer science student participants. After signing consent forms, participants prioritized a backlog of prosocial, antisocial and neutral user stories based on a case description. The process began with a 5-minute instructional video, after which they were asked to begin the prioritization task. One group received a stakeholder map, one group received a set of user personas, one group received both, and the fourth (control) group received neither. All of the materials discussed below are available in our comprehensive replication package (see Section~\ref{sec:availability}).

\begin{figure*}[ht]
\centering
\includegraphics[width=0.90\linewidth]{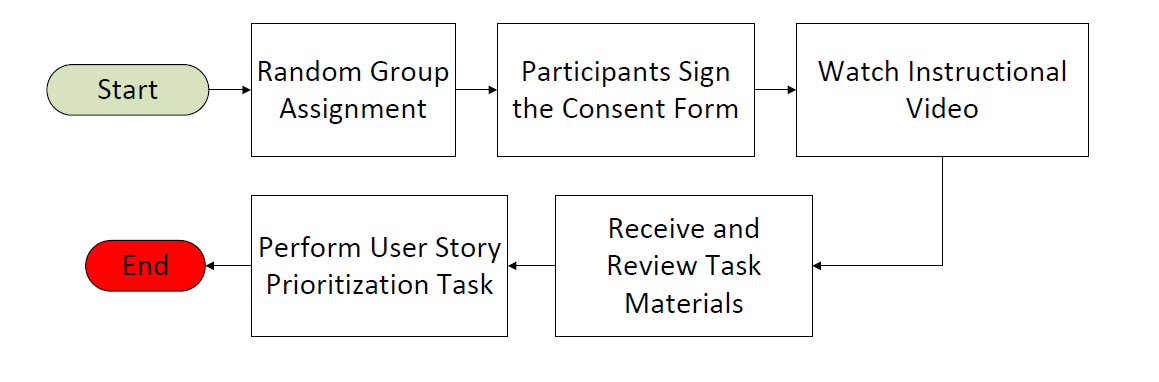}
\caption{Flow chart for the experiment process}
         \label{fig:flow}
\end{figure*}

\subsection{Hypotheses} \label{hyp}
Our hypothesis are as follows:

\renewcommand{\theenumi} {H\arabic{enumi}}
\begin{enumerate}[label=H\arabic*, leftmargin=*]
    \item \begin{enumerate}[label=A:, ref=H\arabic{enumi}A]
        \item Priorities assigned to prosocial user stories are higher when a stakeholder map is presented.
    \end{enumerate}
    \begin{enumerate}[label*=B:, ref=H\arabic{enumi}B]
        \item Priorities assigned to antisocial user stories are lower when a stakeholder map is presented.
    \end{enumerate}
    \item \begin{enumerate}[label=A:, ref=H\arabic{enumi}A]
        \item Priorities assigned to prosocial user stories are higher when a set of persona models is presented.
    \end{enumerate}
    \begin{enumerate}[label*=B:, ref=H\arabic{enumi}B]
        \item Priorities assigned to antisocial user stories are lower when a set of persona models is presented.
    \end{enumerate}
\end{enumerate}
\renewcommand{\theenumi}{\arabic{enumi}}

We chose stakeholder maps and personas because they are frequently used by software professionals and have been shown to affect decision making~\cite{chiarcos2011salience, pooresmaeili2014effect,peters2006affect}. Stakeholder maps not only help decision-makers identify key actors and those impacted by a product but also increase the salience of their needs and perspectives~\cite{power2010stakeholder,newcombe2003client}. \textit{Salience}---the prominence of information---plays a pivotal role in decision-making as it affects how information is perceived and prioritized~\cite{chiarcos2011salience, pooresmaeili2014effect}. Similarly,  persona models deepen understanding of user needs, drive emotional responses, and foster empathy thereby influencing decisions and promoting the consideration of user needs~\cite{peters2006affect}. Persona models and stakeholder maps not only make relevant information more prominent but also act as cognitive enhancers by directing attention to the diverse needs of these groups, thereby improving decision-making outcomes~\cite{beattie1994psychological}.

Moreover, both stakeholder maps and personas can enhancing the visibility of underrepresented groups and implicitly encourage professionals to consider their needs, which is crucial for achieving equitable outcomes.  Simultaneously, personas deepen the understanding of end-user contexts, vital for developing features that are socially beneficial. When both personas and stakeholder maps are used in tandem, they could synergistically enhance software professionals' comprehension of user needs. Our factorial experimental design allows us to investigate the interaction effect when both tools are present.

\subsection{Task Development, Instrumentation, and Pilot Testing} \label{sec:task}

We generated a case description involving a facial recognition system (because of the ethical challenges intrinsic to such systems) at a shopping mall (because we think most software professionals and software engineering students in our region are familiar with shopping malls). We then generated a backlog of user stories representing potential features of the facial recognition system. 

We conducted a pilot study comprising four rounds of:
\begin{itemize}
    \item six individuals from diverse cultural background categorized each story as prosocial, neutral, or antisocial;
    \item for each story with less than 100\% agreement, we revised, removed, or replaced the story.
\end{itemize}
After four rounds we reached 100\% agreement on 27 user stories: 9 prosocial (features that would benefit social justice); 9 antisocial (features that would harm social justice), and 9 neutral (with no obvious implications for social justice). The number of stories, 27, was carefully chosen to present a realistic workload without overwhelming the participants or degrading interest. We included neutral stories because a backlog with only pro- and anti-social stories, and no middle ground, would be unrealistic.

In real life, of course, stories could occupy a continuous spectrum or have simultaneous pro- and anti-social effects. The pilot process may have eliminated stories that were borderline or had mixed effects, and intentionally so. Including borderline or mixed-effect stories would decrease objectivity and reliability (by hindering consensus in the pilot process) while potentially making our statistical analysis intractable. 

\begin{figure*}[ht]
\centering
\includegraphics[width=0.85\linewidth]{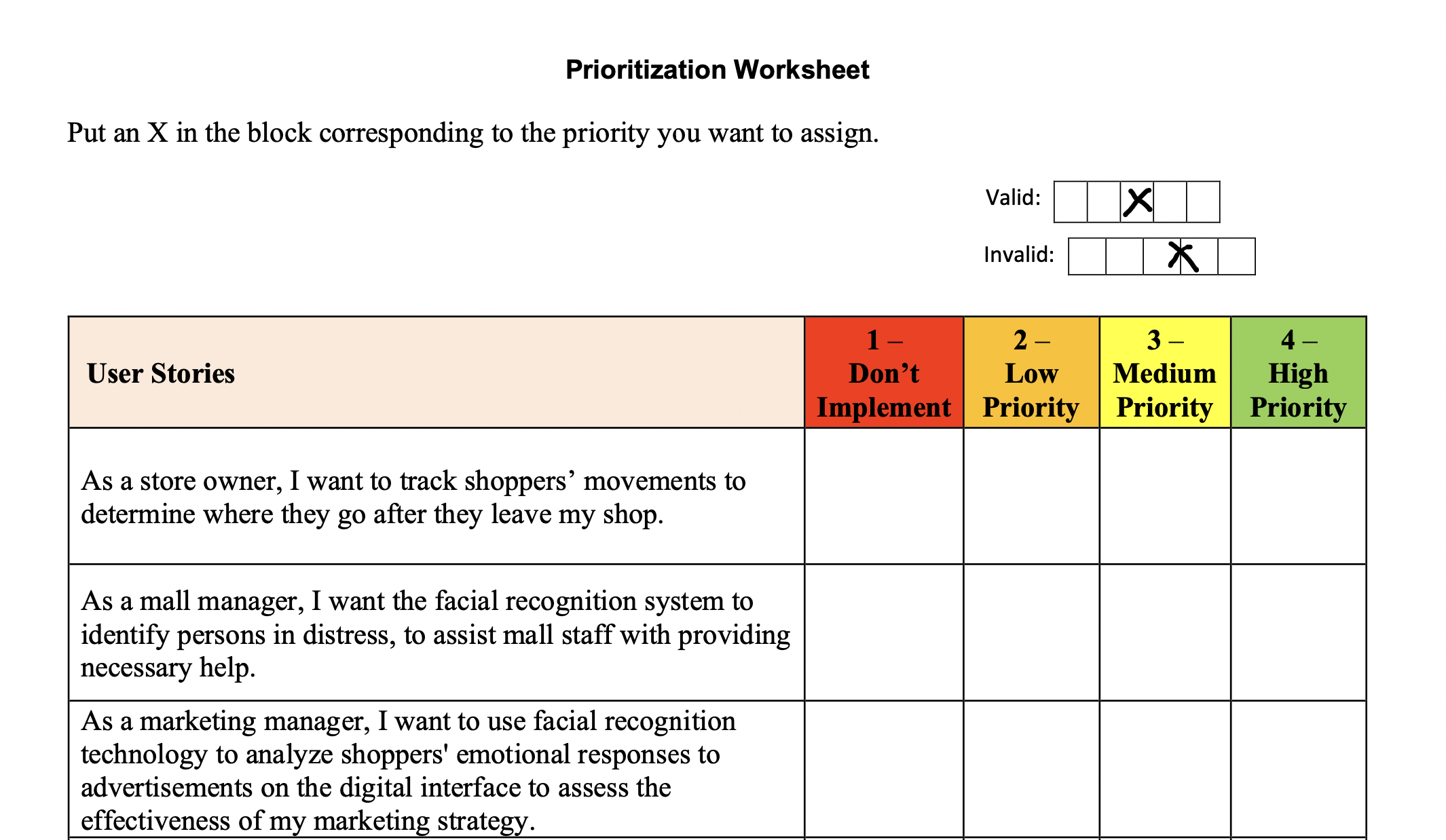}
\caption{Sample Prioritization Worksheet}
         \label{fig:worksheet}
\end{figure*}

We then implemented the experimental task on paper to encourage more deliberate and thoughtful decision-making without distractions. All participants were given the shopping mall / facial recognition case to provide context. Participants in the experiment groups were given a stakeholder map and/or personas for the prioritization task and instructed to consider the stakeholder map, personas, or both, in assessing the potential impact of each feature on various stakeholders or persona models as they carry out the user story prioritization. Participants in the control group received only the case with neither intervention. 
Participants were asked to assign each story one of four priorities---1 (don't implement), 2 (low priority), 3 (medium priority), 4 (high priority)---by ticking the appropriate level (see Fig.~\ref{fig:worksheet}). 
The list of user stories was also deliberately rearranged and mixed up for each participant to reduce selection bias and also ensure that the order of presentation did not influence their prioritization decisions. This rearrangement involved manually copying and pasting the user stories in different sequences.

\begin{figure}[ht]
\centering
\includegraphics[width=0.99\linewidth]{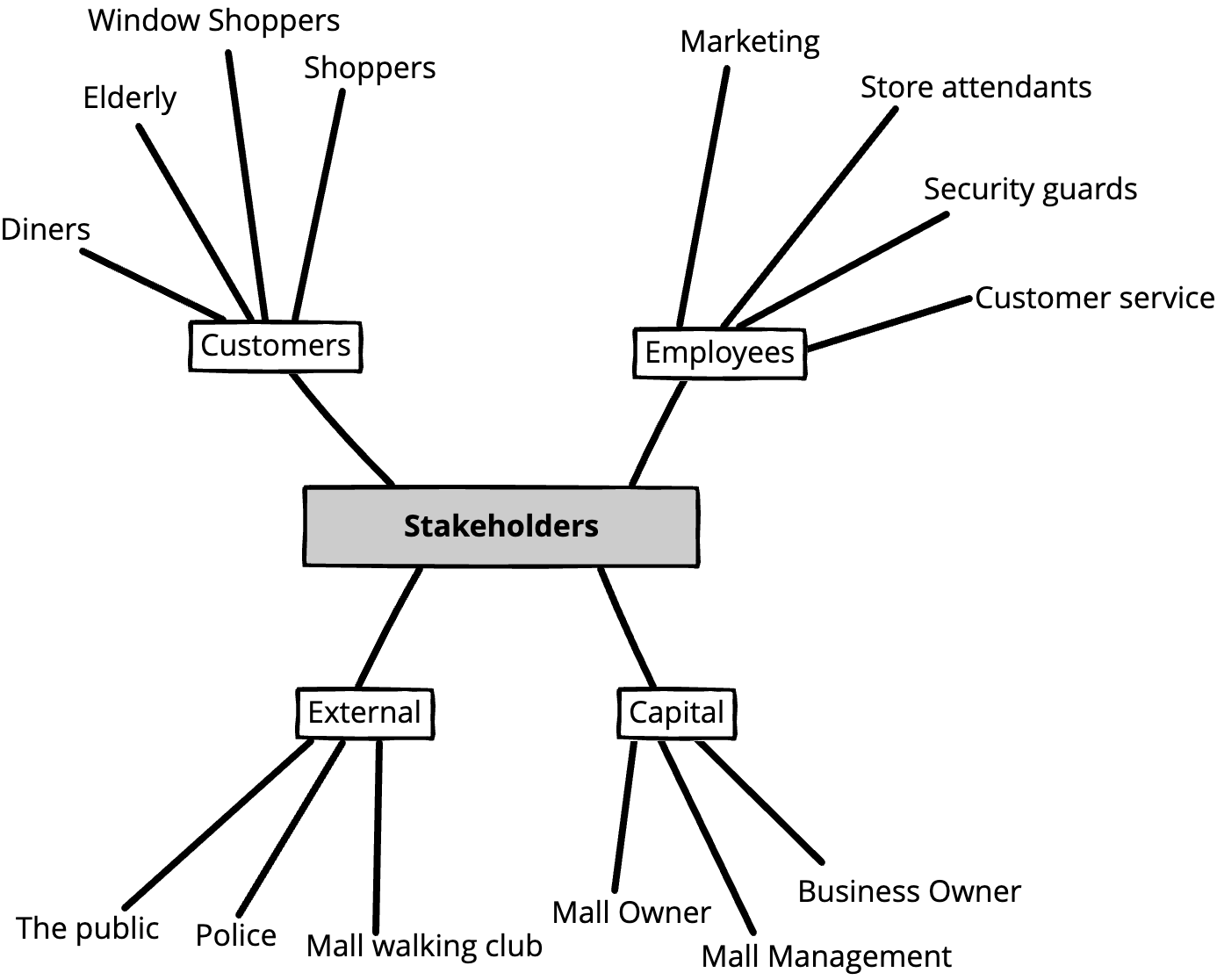}
\caption{Sample stakeholder map given to the participants}
         \label{fig:map}
\end{figure}

\subsection{Recruitment} \label{rq}

We used the GPower~\cite{erdfelder1996gpower} (version 3.1) to estimate the required sample size for this study. Assuming a medium effect size of 0.4, 95\% confidence level, and an alpha level 0.05 yielded an estimated sample size of 104.

Participants were recruited directly from an undergraduate computer science course at the authors' university. Participating in the study was one of two assignments students could complete for 2\% of their term grade. The options were described in the syllabus and participants were recruited through announcements made through the learning management system. The two assignment options were clearly explained, and the alternative assignment was less onerous than participating in the experiment so as to avoid coercion. Of the 92 students enrolled in the course, 81 attended the experiment session and 79 completed the experiment task (see Table~\ref{tab:summary_stat}).

\subsection{Protocol} \label{protocol}
We used Excel's \textsf{RAND} function to assign each invited participant to one of four groups: control (no stakeholder map or personas), stakeholder map only, personas only, or both stakeholder map and personas---see Table~\ref{tab:summary_stat}. Each group was assigned to one of four different rooms on campus, which were selected for their similar layouts and setups to maintain consistency in the experimental environment. We then emailed their room assignments (and reminded them about the right to withdraw and opportunity to complete the alternative assignment).

Each experiment group had two invigilators: the first and second authors and six computer science graduate students. Before the experiment, all invigilators received a detailed briefing from the first author. The briefing focused on the experiment's objectives and procedures, providing step-by-step guidance for the experiment day. Invigilators also received a written experimental protocol.

The experiment was conducted simultaneously with all participants on November 23\textsuperscript{rd}, 2023. All participants were asked to sign a consent form and complete a demographic questionnaire including questions about age, gender, and nationality (see Table~\ref{tab:participant_summary}). 
Participants were then asked to watch an instructional video, which had been pre-recorded by the first author. The video describes the study task and materials.  After watching the video, participants were then asked review the task materials and prioritize a backlog of user stories for a mall's digital screen display with facial recognition software, as explained in Section~\ref{sec:task}. This task took approximately 50 minutes to complete.

\begin{table}[t]
\centering
\caption{Group Participant Counts }
\label{tab:summary_stat}
\begin{tabular}{lcccc}
\toprule
{} &  Invited &  No show  &  Withdrew & Participants\\ 
\midrule
Control      &      23 &        3 &        0 & 20\\
Stakeholder map &        22 &        3 &        0 & 19\\
Persona        &         23 &        5 &        2 & 16\\
S. map * Persona &        24 &        0 &        0 & 24\\
\midrule
Total & 92 & 11 & 2 & 79\\
\bottomrule
\end{tabular}
\end{table}

\begin{table}[t]
\begin{threeparttable}[b]
\setlength{\tabcolsep}{2pt} 
\renewcommand{\arraystretch}{1.1} 
\centering
\caption{Participant Summary}
\label{tab:participant_summary}
\begin{tabularx}{\linewidth}{lXrr}
\toprule
{Dimension} & {Attribute} & {n} & {\%} \\
\midrule
\multirow{3}{*}{Gender} & Female & 15 & 18.99 \\
                        & Male & 59 & 74.68 \\
                        & Prefer not to say & 5 & 6.33 \\
\midrule
\multirow{4}{*}{Age} & 15-20 & 22 & 27.85 \\
                     & 21-26 & 50 & 63.29 \\
                     & 27-32 & 2 & 2.53 \\
                     & Prefer not to say & 4 & 5.06 \\
\midrule
\multirow{5}{*}{Country} & Canada & 29 & 36.71 \\
                         & China & 14 & 17.72 \\
                         & India & 12 & 15.19 \\
                         & Others (Afghanistan, Bangladesh, Barbados, Colombia, Egypt, Israel, Lebanon, Oman, Pakistan, Palestine, Portugal, Saudi Arabia, USA) & 20 & 25.35 \\
                         & Prefer not to say & 4 & 5.06 \\
\bottomrule
\end{tabularx}%
\begin{tablenotes}
       \item *Some columns do not total 100\% due to rounding.
\end{tablenotes}
\end{threeparttable}
\end{table}

\subsection{Data Analysis} \label{ssec:analysis}

All collected data were double-entered into a spreadsheet by the first and second authors. Disagreements were found for 95 rows (4\% of the data) and corrected by consensus by double checking the original data with both authors present. 

Each row is a data point where a participant has prioritized a particular story type. Columns include participant ID, user story ID, story type, treatment group, priority assigned, gender, age, and country. Story type is a categorical variable with three possible levels: prosocial, neutral and antisocial. Treatment group is a categorical variable with four possible values: persona, stakeholder map, both, and control. Lastly, the prioritization level is an ordinal variable with values from 1 to 4.

Our analysis began by examining whether our dataset satisfies assumptions for Cumulative Link Mixed Model (CLMM), an extension of Ordinal Logistic Regression (OLR) (see section~\ref{ssec:diag}). CLMM is used to predict an ordinal dependent variable given one or more independent variables, accounting for fixed and random effects. CLMM is necessary because each participant prioritizes multiple stories, making the observations not wholly independent. 

Next, we split the dataset and fit two CLMM models, one predicting the priorities assigned to prosocial stories; the other predicting the categories assigned to antisocial stories. We took a dual model approach for two reasons. First, our hypotheses focus on the interactions between the type of user story (prosocial, neutral, or antisocial) and the intervention groups (stakeholder map, personas, both, or neither). A single model would be more complicated and difficult to understand because it includes main effects (which are irrelevant to our hypotheses); neutral stories (also irrelevant to our hypotheses), and many more coefficients. Second, fitting a single model would lead to violations of CLMM's proportional odds assumption. Thus, we tested our hypotheses using two separate CLMM models.

For both models, the dependent variable is the prioritization level, which is ordinal. We recoded the categorical variable of the intervention group to binary variables~\cite{hardy2004incorporating}, which allowed us to set the reference category by the order of independent binary variables. We set the control group as the reference category; that is, the category of the variable to which the other levels of the categorical variable are compared. 

We fit the model using the \textsf{ordinal} package~\cite{christensen2019ordinal} for the R programming language (Version 4.1.2)~\cite{RCoreTeam2021}.

Recall that we hypothesized that the stakeholder map and personas would lead to higher prioritization of prosocial user stories and lower prioritization of antisocial user stories. A significant negative coefficient for any of the predictors `persona', `stakeholdermap', and `stakeholdermap $*$ persona' with respect to the story type would indicate that the use of personas or stakeholder map or both effectively influences participants to prioritize antisocial stories lower or prosocial stories higher. Predictors with p-values less than 0.025 will be considered statistically significant.

Detailed analysis scripts are available in our comprehensive replication package (see Section~\ref{sec:availability})

\section{Results} \label{sec:synthesis_results}

In this section, we report data diagnostics and hypothesis tests, and visualize our main results. 

\subsection{Data Diagnostics} \label{ssec:diag}

CLMM makes four main assumptions:

\begin{enumerate}

    \item The dependent variable is ordinal. Our dependent variable, Prioritization\_Level, is in fact ordinal.
    
    \item The independent variable is categorical. We recoded our independent variables to make them binary~\cite{hardy2004incorporating}, which satisfies this assumption.
    
    \item No multicollinearity among independent variables. We checked for multicollinearity with the \textsf{vif} function from the \textsf{car} package~\cite{fox2012package}. All variance inflation factors were between 1.42 and 1.53, well below the threshold of 5~\cite{akinwande2015variance}.
    
    \item The effect of each independent variable is consistent across different thresholds of the ordinal dependent variable (Proportional Odds). We used the function \textsf{nominal\_test} from the \textsf{ordinal} package to check for proportional odds~\cite{christensen2019ordinal}. The data was split by story type as described in Section~\ref{ssec:analysis}. The results show high p-values ($p > 0.05$) for `stakeholder map' ($p = 0.23, 0.12$), `persona' ($p = 0.69,0.58$) and their interaction ($p = 0.072,0.068$), indicating that the Proportional Odds assumption is met.

\end{enumerate}

\begin{table*}
    \centering
    \setlength{\tabcolsep}{15pt}
    \caption{Cumulative Link Mixed Model (CLMM) Results for Prosocial User Stories Prioritization}
    \label{tab:tab_CLMM-Prosocial}
\begin{tabular}{l l c c c c}
\toprule
\textbf{Random Intercept for:}                & \multicolumn{4}{l}{Participant.ID } \\

\textbf{Intercept Variance}              & \multicolumn{4}{l}{$0.1475$} \\
\textbf{Intercept Std.Dev.}              & \multicolumn{4}{l}{$0.3840$} \\
\textbf{Number of Participants}      & \multicolumn{4}{l}{$79$} \\
\midrule

& \textbf{Coefficients} & \textbf{Std. Error} & \textbf{z value} & \textbf{P \textgreater{} \textbar z\textbar} & \textbf{odds ratio}\\
\midrule


\textbf{Reference Category: Control} \\ 

\textbf{Stakeholdermap}  & $0.3124$ & $0.2344$ & $1.333$ & $0.1830$ & $1.3666$\\

\textbf{Persona}         & $0.1645$ & $0.2442$ & $0.674$ & $0.5010$ & $1.1787$\\

\textbf{Both}           & $-0.0585$ & $0.2163$ & $-0.270$ & $0.7870$ & $0.9432$ \\


\bottomrule
\end{tabular}

\end{table*}

\begin{table*}
    \centering
    \setlength{\tabcolsep}{15pt}
    \caption{Cumulative Link Mixed Model (CLMM) Results for Antisocial User Stories Prioritization}
    \label{tab:tab_CLMM-Antisocial}
\begin{tabular}{l c c c l c}
\toprule
\textbf{Random Intercept for:}   & \multicolumn{4}{l}{Participant.ID} \\

\textbf{Intercept Variance}     & \multicolumn{4}{l}{$0.5759$} \\
\textbf{Intercept Std.Dev.}     & \multicolumn{4}{l}{$0.7589$} \\
\textbf{Number of Participants} & \multicolumn{4}{l}{$79$} \\
\midrule
& \textbf{Coefficients} & \textbf{Std. Error} & \textbf{z value} & \textbf{P \textgreater{} \textbar z\textbar} & \textbf{odds ratio} \\ 
\midrule



\textbf{Reference Category: Control} \\

\textbf{Stakeholdermap}  & $-0.2059$ & $0.3152$ & $-0.653$ & $0.5135$ & $0.8139$\\

\textbf{Persona}         & $-0.8672$ & $0.3314$ & $-2.617$ & $0.0088$ ** & $0.4201$\\

\textbf{Both}            & $-0.5258$ & $0.2966$ & $-1.773$ & $0.0762$ & $0.5911$ \\

\bottomrule
\end{tabular}



\end{table*}

\subsection{Hypothesis Tests} \label{ssec:hypTests}

We test each hypothesis below using CLMM regression.

\textbf{Hypothesis H1A: Priorities assigned to prosocial user stories are higher when a stakeholder map is presented.}
With a positive coefficient of $0.3124$ and a p-value $0.183$, we fail to reject the null hypothesis (see Table \ref{tab:tab_CLMM-Prosocial}). The effect of the stakeholder map on prioritization of prosocial stories is not statistically significant.

\textbf{Hypothesis H1B: Priorities assigned to antisocial user stories are lower when a stakeholder map is presented.}
With a coefficient of $-0.2059$ with a p-value of $0.5135$, we fail to reject the null hypothesis (see Table \ref{tab:tab_CLMM-Antisocial}). The effect of the stakeholder map on prioritization of antisocial stories is not statistically significant.

\textbf{Hypothesis H2A: Priorities assigned to prosocial user stories are higher when a set of persona models is presented.}
With a positive coefficient of $0.1645$ and p-value $0.501$, we fail to reject the null hypothesis (see Table \ref{tab:tab_CLMM-Prosocial}). The effect of personas on prioritization of prosocial user stories is not statistically significant.

\textbf{Hypothesis H2B: Priorities assigned to antisocial user stories are lower when a set of persona models is presented.}
With a significant negative coefficient of $-0.8672$ $(p = 0.009)$, we reject the null hypothesis. The effect of personas on prioritization of antisocial stories is statistically significant (see Table \ref{tab:tab_CLMM-Antisocial}). The corresponding odds ratio $(0.42)$ indicates that the presence of personas reduces the odds of assigning a higher prioritization level to antisocial user stories by $58\%$.

\textbf{Interaction effects.} If there were a synergistic interaction between the treatments (i.e., stakeholder maps and personas are more effective together than individually) then absolute value of the coefficients (and odds ratios) of the experimental group receiving both treatments should exceed the absolute value of the coefficients (and odds ratios) of the experimental groups that received either treatment alone. This is not the case, so there is no evidence of synergistic interaction. 

\subsection{Data Visualization} \label{ssec:dataViz}
\begin{figure*}[ht]
    
    \centering
    \includegraphics[width=0.95\linewidth]{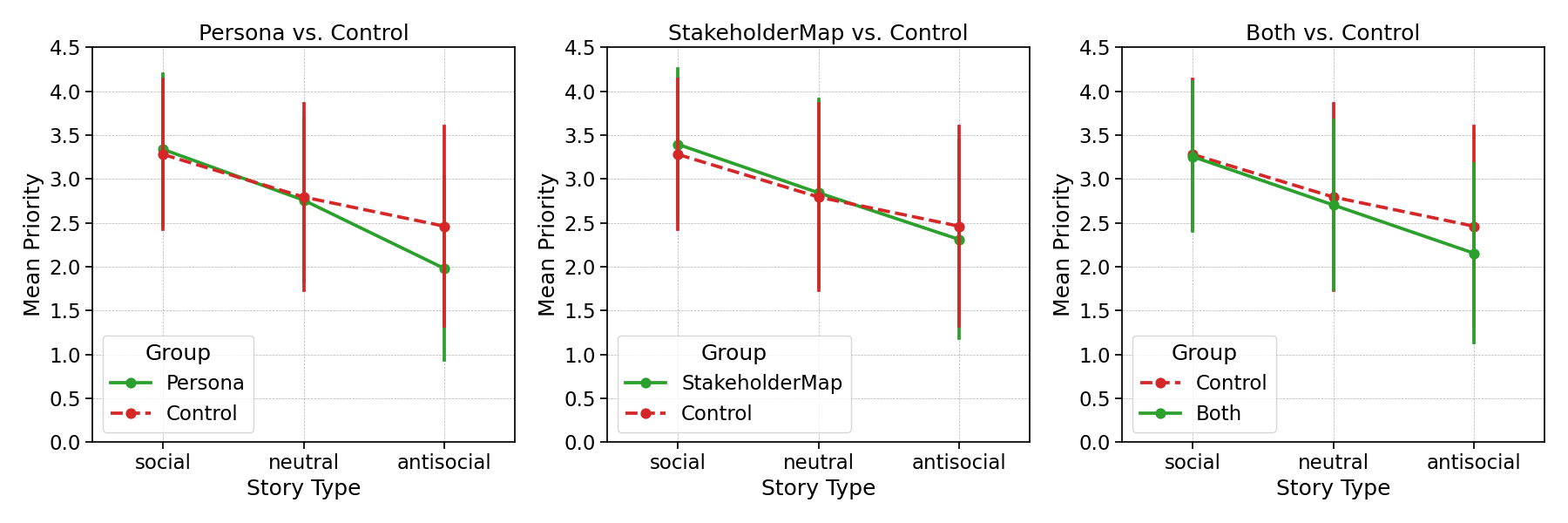}
    \caption{Interaction Plots}
    \label{fig:interactionPlot}
    
\end{figure*}

\begin{table}[t]
\centering
\caption{Mean Prioritization Scores per Group and User Story Type}
\label{tab:meanscorespergroup}
\begin{tabular}{lcccc}
\toprule
 & \textbf{antisocial}  & \textbf{neutral} &\textbf{social} & \textbf{n} \\
\midrule
\textbf{Stakeholder map} & $2.30$ & $2.84$ & $3.39$ & $19$ \\
\textbf{Persona} & $1.98$ & $2.74$ & $3.34$ & $16$ \\
\textbf{Both} & $2.15$ & $2.70$ & $3.28$ & $24$ \\
\textbf{Control} & $2.46$ & $2.79$ & $3.28$ & $20$ \\
\bottomrule
\end{tabular}
\end{table}

Fig.~\ref{fig:interactionPlot} and Table~\ref{tab:meanscorespergroup} show mean prioritization scores across different intervention groups and user story types. The negative slopes of the lines indicate that participants in all groups prioritized prosocial stories higher than neutral stories and neutral stories higher than antisocial stories, on average.

The distances between the dashed red lines and solid green lines correspond to our hypothesis tests. In the leftmost plot, we see that participants gave similar average priorities to prosocial and neutral stories regardless of whether they received persona models. However, participants who received persona models assigned significantly lower priorities to antisocial user stories.

The middle plot shows how participants who received the stakeholder map gave slightly higher priorities to prosocial stories, and slightly lower scores to antisocial stories compared to the control group, but neither difference is statistically significant. 

The rightmost plot shows how participants who received both the stakeholder map and personas gave somewhat lower scores to antisocial user stories, but this difference is not significant. Moreover, because the ``both'' group displays a smaller effect than the persona-only group, we see no visual evidence of synergistic interaction between the treatments.  

\section{Discussion} \label{sec:discussion}

We hypothesized that exposure to a stakeholder map and persona models would lead participants to prioritize prosocial user stories higher and antisocial user stories lower. We found that the effect of personas is imbalanced: personas significantly lower priorities of antisocial stories but do not increase priorities of prosocial stories. 

Stakeholder maps appear to display a more balanced pattern (boosting prosocial story priorities while dampening antisocial story priorities); however, these effects are not statistically significant, either because the pattern is just random noise or because the effect is too small to detect with the study's present sample size. 

Similarly, we found no evidence of synergistic (or antagonistic) interaction between treatments. Again, this may be because the treatments do not interact or because the effect of stakeholder maps is too small to detect with a study of this size. 

The fact that personas have a stronger effect is consistent with our proposed generative mechanism. Personas should affect prioritization by helping professionals empathize with users. More diverse personas, as used in this study, should help professionals empathize with a greater variety of users. In this sense, personas, with their pictures and details, are a more powerful trigger than stakeholder maps. 

\subsection{Implications} \label{sec:implications}

A recent scoping review found a lack of studies evaluating sustainability-improving interventions~\cite{Sean_sust}. Meanwhile, Becker~\cite{becker2023insolvent} emphasized the importance of enacting meaningful change in computing to address pressing sustainability issues. These works inspired the study described in this paper, which investigates specific interventions that could improve social sustainability in software development.

Based on the results reported above, we recommend that software teams consider incorporating personas into their software development process to promote socially sustainable development. Although personas only affected the prioritization of antisocial stories in our experiment, practically speaking, priorities are relative, so it may not matter much if an intervention boosts prosocial priorities or dampens antisocial priorities or a bit of both.

For researchers, this study contributes to the growing body of knowledge regarding the intersection of software engineering and social sustainability. To date, there is a lack of empirical (particularly experimental) validation of specific social sustainability interventions, making our findings particularly noteworthy. 

Furthermore, we suspect that the lack of experiments evaluating social sustainability interventions is due, in part, to the difficulty of assessing these interventions. We hope that other researchers will adopt our backlog-prioritization approach to assessing social sustainability interventions and provide all of our materials for this purpose (see \nameref{sec:availability}). A common assessment approach is more efficient and should improve replicability and validity (as different research teams refine and evolve one approach instead of re-inventing new ones). 

For educators, our results support teaching persona modeling in the software engineering curriculum. Furthermore, we encourage educators to teach sustainable development in general and social sustainability in particular throughout computer science and engineering curricula. Simply promoting sustainability awareness and normalizing discussions thereof is half the battle.

\subsection{Limitations} \label{sec:limitations}
In this section we address standard quality criteria for experiments with human participants~\cite{ralph2020empirical}. Like many randomized controlled experiments, this study sacrifices generalizability to maximize internal validity.

Internal validity is high because: (1) we used a well-understood, widely-used, fully-crossed, two-factor between subjects design in which participants were randomly assigned to experimental groups; (2) participants were unaware of the specific research focus (reducing social desirability bias); (3) each group was in its own room, reducing social threats to internal validity; (4) all participants were provided with the same prioritization worksheet and placed in similar experiment rooms with uniform experimental setups. 

Data were double-entered to improve reliability and the high level of agreement achieved demonstrates good reliability. Conclusion validity is similarly high as we used a sophisticated regression technique (CLMM) that maximizes statistical power given constraints imposed by data types and distributions. We systematically identified all the assumptions of CLMM and verified that they are met). However, our sample size of 79 fell short of the suggested sample size of 104, inflating the probability of a type II error.  

However, strong internal validity, reliability and conclusion validity come at the cost of poor external validity. We used a convenience sample of participants. While multicultural, the participants are all students in the same university so they may have other unknown similarities. Participants were students, not professionals, further threatening external validity (see~\cite{salman2015students,falessi2018empirical}). We used a single case (shopping mall facial recognition), a single stakeholder map, and a specific set of personas. Determining how sensitive our results are to all these specific details is not possible with this kind of study. A different sample of participants analyzing a different case using different task materials may act differently. While serious, these threats to internal validity are endemic to experiments with human participants; the whole point of an experiment is to maximize internal validity, not external validity~\cite{stol2018abc}.

Meanwhile, the construct validity of our study is nuanced. The priorities assigned by participants to stories are measured directly, not reflectively, so a psychometric approach to construct validity (e.g. convergent and discriminant validity) does not apply. We consulted with several industry experts to to ensure the priority levels reflect contemporary agile practice, and the prioritization categories were clearly explained to all participants in the task material and in the video shown at the beginning of the study. However, the reader is justified in wondering to what extent these priority ratings relate to social sustainability. In the broadest sense, pro- and antisocial simply mean good or bad for society, but there are no authoritative lists of what exactly are good or bad for society, and reasonable people both within and across cultures may disagree on whether a specific feature is good or bad for society. 

The validity of our prioritization-based measurement approach therefore rests on our categorization of stories into prosocial, neutral, and antisocial. As explained in Section~\ref{sec:task}, we conducted a series of pilot studies to ensure high agreement about the nature of each story. This approach is not perfect. It assumes that participants in the pilot study can estimate the social impact of the proposed user stories given the source materials and that the user stories can be neatly classified. Not every feature of every real software product can be neatly binned into prosocial, antisocial, or neutral, so maximizing agreement likely excludes more ambiguous stories. Furthermore, the pilot participants are not a representative sample of all of humanity.  Nevertheless, the perfect agreement reached through the pilot process suggests good face validity. Furthermore, the fact that control group participants on average prioritized prosocial stories significantly higher than neutral stories and neutral stories significantly higher than antisocial stories suggests good predictive validity. In summary, then, since convergent and discriminant validity do not apply, we took reasonable steps to assess face and predictive validity. However, the imperfect nature of our pilot studies and oversimplification of features' social effects threaten construct validity in unquantifiable ways. Personas may be less effective at promoting social sustainability in projects full of controversial features that could be perceived as either pro- or antisocial. 

\subsection{Proposed Research Methodology for Social Sustainability Experiments}
The lack of experimental evaluations of social sustainability-promoting interventions may be due in part to the methodological challenges of constructing such studies. Studies of software development processes (and modifications thereof) resist experimental closure~\cite{ralph2010fundamentals}. That is, it is not possible to recruit hundreds of real software teams, randomly assign them to multiple groups, get one group to use some intervention while another does not, then measure the sustainability of the systems they produce. 

We therefore designed an experimental protocol for assessing social sustainability interventions to overcome four methodological challenges:

\begin{enumerate}
    \item Expressed attitudes do not generally correspond to revealed attitudes~\cite{fishbein1975belief}, due to social desirability bias among other psychological factors~\cite{fisher1993social}. The experimental task therefore should involve observable actions beyond expressing attitudes toward sustainability.

    \item Most software professionals should be able to complete the experimental task. Therefore, we need a familiar, everyday context that does not require specialized domain knowledge (e.g. a web browser, not air traffic control software). 
    
    \item The methodology should be implementable as either an in-person, lab-based task (suitable for student participants) or a remote, web-based task (suitable for recruiting a geographically diverse sample of professionals).
    
    \item The methodology should be easily adaptable to many different social sustainability promoting interventions, both to improve comparability across studies and to reduce effort associated with study design.
\end{enumerate}

With these goals in mind, we formulated the method described in Section~\ref{sec:method}. Developing the task materials took many iterations---significantly more effort than previous experiments with human participants we have conducted---but the resulting method appears to achieve all of our goals. 

The task of prioritizing a backlog of user stories provides a good compromise. It is something software professionals do in the real world~\cite{sedano2019product} that is easy to simulate and can give us insight into propensities (not just attitudes). When prioritizing the stories, participants consider many factors beyond social sustainability, so the tasks give us clues as to how much social sustainability considerations might affect participants' behaviour. 

Meanwhile, we chose a facial recognition system because it is so ethically fraught that it is easy to generate clearly antisocial features. We chose a shopping mall setting because most people are familiar with shopping malls---shopping malls are ubiquitous throughout developed and middle-income countries and while they are concentrated in urban areas, so are software professionals. We pilot-tested our user stories with a culturally diverse group to reduce the likelihood that our pro- and anti-social categorization is culturally-relative.

While we implemented the study on paper, it is conducive to an online presentation. An interface for collecting story priorities would be trivial to create with any common questionnaire survey software. The only important difference is the ability to spread materials across a table. Interventions involving copious information may be more difficult to implement online.

In summary, we advance a novel approach for assessing social sustainability interventions in which experimental or simulation participants prioritize a backlog of prosocial, antisocial, and neutral user stories, and researchers compare mean priority ratings. We provide an easy-to-follow experimental protocol (Section~\ref{sec:method}) and materials (see~\nameref{sec:availability}) for other researchers to repeat the approach. Finding a reasonable way of assessing the impact of sustainability interventions in lab studies is an important step for facilitating future research in this area (next). 

\subsection{Future Work} \label{sec:future}
Several avenues of future work are evident, including:
\begin{itemize}
    \item Replicating the present study with professionals
    \item Conducting methodologically similar studies testing different interventions (e.g. exposure to postgraduate education in ethics and sustainability; participatory design).
    \item Refining our evaluation methodology to include: (1) ethically murkier stories with mixtures of pro- and anti-social effects; (2) greater variety of pro- and anti-social effects.
\end{itemize}

More studies proposing and rigorously evaluating interventions for promoting environmental, economic, and technical sustainability are also sorely needed. 

\section{Conclusion} \label{sec:conclusion}

Sustainability is not just about building technically robust or energy-efficient products. Sustainability is also about infusing human values, ethics, and social responsibility into software products. The human and social aspects of software engineering are often neglected; yet it is increasingly clear that social sustainability is crucial for building software that makes the world better instead of worse. Recent studies have called on software engineering researchers to propose and validate more sustainability-promoting interventions (and write fewer position papers)~\cite{Sean_sust, moises2023social}. Our study has shown the possibilities of validating and assessing these kind of tools, starting with a controlled lab setting with student participants. 

This paper makes two main contributions: 
\begin{enumerate}
    \item The first empirical evidence that a common software development practice (persona modeling) can affect social sustainability (in this case, by discouraging the selection and prioritization of antisocial features).
    \item We provide a replicable, adaptable, extendable experimental approach for evaluating social sustainability interventions in lab- or web-based settings. 
\end{enumerate}

In summary, our findings suggest that technology companies and software professionals who value social sustainability should try user persona modeling. While we cannot say for sure that our lab study with students will generalize to any particular organizational context, persona modeling has no known risks or drawbacks.  

We hope this study inspires other researchers to create and experimentally assess other sustainability-promoting interventions.

\section*{Acknowledgments}
This project was supported by National Sciences and Engineering Research Council of Canada Discovery Grant RGPIN-2020-05001, and Discovery Accelerator Supplement RGPAS-2020-00081.

\section*{Data Availability} \label{sec:availability}
We provide a detailed replication package including our pilot study materials, experiment task materials, instructional videos, deidentified dataset, data diagnostics and data analysis script (\url{https://doi.org/10.5281/zenodo.10827123}).

\ifCLASSOPTIONcaptionsoff
  \newpage
\fi

\bibliographystyle{IEEEtran}

\balance

\bibliography{bib.bib}

\end{document}